\documentclass[11pt]{article}
\usepackage{amssymb,cite}
\makeatletter
\@addtoreset{equation}{section}
\makeatother

\def\ben{\begin{equation}}
\def\een{\end{equation}}

  \let\n=\nu

\let\C=\Chi

\def\nn{\nonumber} \def\bd{\begin{document}} \def\ed{\end{document}}
\def\ds{\documentstyle} \let\fr=\frac \let\bl=\bigl \let\br=\bigr
\let\Br=\Bigr \let\Bl=\Bigl
\let\bm=\bibitem
\let\na=\nabla
\let\pa=\partial \let\ov=\overline
\newcommand{\be}{\begin{equation}}
\newcommand{\ee}{\end{equation}}
\def\ba{\begin{array}}
\def\ea{\end{array}}
\def\ft#1#2{{\textstyle{{\scriptstyle #1}\over {\scriptstyle #2}}}}
\def\fft#1#2{{#1 \over #2}}
\def\del{\partial}
\def\vp{\varphi}
\def\sst#1{{\scriptscriptstyle #1}}
\def\oneone{\rlap 1\mkern4mu{\rm l}}
\def\td{\tilde}
\def\wtd{\widetilde}
\def\ie{\rm i.e.\ }
\def\dalemb#1#2{{\vbox{\hrule height .#2pt
        \hbox{\vrule width.#2pt height#1pt \kern#1pt
                \vrule width.#2pt}
        \hrule height.#2pt}}}
\def\square{\mathord{\dalemb{6.8}{7}\hbox{\hskip1pt}}}
\newcommand{\ho}[1]{$\, ^{#1}$}
\newcommand{\hoch}[1]{$\, ^{#1}$}
\newcommand{\bea}{\begin{eqnarray}}
\newcommand{\eea}{\end{eqnarray}}
\newcommand{\ra}{\rightarrow}
\newcommand{\lra}{\longrightarrow}
\newcommand{\Lra}{\Leftrightarrow}
\newcommand{\ap}{\alpha^\prime}
\newcommand{\bp}{\tilde \beta^\prime}
\newcommand{\tr}{{\rm tr} }
\newcommand{\Tr}{{\rm Tr} }
\def\0{{\sst{(0)}}}
\def\1{{\sst{(1)}}}
\def\2{{\sst{(2)}}}
\def\3{{\sst{(3)}}}
\def\4{{\sst{(4)}}}
\def\5{{\sst{(5)}}}
\def\6{{\sst{(6)}}}
\def\7{{\sst{(7)}}}
\def\8{{\sst{(8)}}}
\def\n{{\sst{(n)}}}
\def\cA{{{\cal A}}}
\def\cB{{{\cal B}}}
\def\cF{{{\cal F}}}
\def\cH{{{\cal H}}}
\def\tV{\widetilde V}
\def\tW{\widetilde W}
\def\tH{\widetilde H}
\def\tE{\widetilde E}
\def\tF{\widetilde F}
\def\tA{\widetilde A}
\def\im{{{\rm i}}}
\def\tY{{{\wtd Y}}}
\def\ep{{\epsilon}}
\def\vep{{\varepsilon}}
\def\R{\rlap{\rm I}\mkern3mu{\rm R}}
\def\bD{{{\bar D}}}
\def\R{\rlap{\rm I}\mkern3mu{\rm R}}
\def\bD{{{\bar D}}}
\def\R{{{\mathbb R}}}
\def\C{{{\mathbb C}}}
\def\H{{{\mathbb H}}}
\def\CP{{{\mathbb C}{\mathbb P}}}
\def\RP{{{\mathbb R}{\mathbb P}}}
\def\Z{{{\mathbb Z}}}
\def\bA{{{\mathbb A}}}
\def\bB{{{\mathbb B}}}
\def\bC{{{\mathbb C}}}
\def\bD{{{\mathbb D}}}
\def\bE{{{\mathbb E}}}
\def\bZ{{{\mathbb Z}}}
\def\Re{{{\mathfrak{Re}}}}
\def\Im{{{\mathfrak{Im}}}}
\def\cosec{{\,\hbox{cosec}\,}}
\def\Gm{{\Gamma_{\!\! -}}}
\def\Gp{{\Gamma_{\!\! +}}}
\def\stan{{standard }}
\def\nonstan{{supernumerary }}
\def\FF2{{ {}_{\sst 2}F_{\sst 1} }}
\def\FFF{{ {}_{\sst 3}F_{\sst 2} }}
\def\const{\rm constant}

\newcommand{\auth}{Muraari Vasudevan$^*$, Kory A. Stevens$^\dagger$ and Don N. Page$^{\ddagger}$}

\begin{document}
\begin{flushright}

Alberta Thy 15-04\\

July\  2004
\end{flushright}

\vspace{10pt}

\begin{center}

{\large {\bf Particle Motion and Scalar Field Propagation in Myers-Perry Black Hole Spacetimes in All Dimensions
            }}

\vspace{20pt}
\auth

\vspace{20pt}

\vspace{10pt}

{\it Theoretical Physics Institute,\\ University of Alberta,
Edmonton, Alberta  T6G 2J1, Canada}

{\it
  \medskip  {$^*$\rm E-mail: \texttt{mvasudev@phys.ualberta.ca}}
  }

{\it
  \medskip  {$^\dagger$\rm E-mail: \texttt{kstevens@phys.ualberta.ca}}
  }

{\it
  \medskip  {$^{\dagger \dagger}$\rm E-mail: \texttt{don@phys.ualberta.ca}}
  }


\vspace{40pt}

\underline{ABSTRACT}
\end{center}

We study separability of the Hamilton-Jacobi and massive Klein-Gordon equations
in the general Myers-Perry black hole background in all dimensions. Complete separation of both equations is
carried out in cases when there are two sets of equal black hole rotation parameters, which significantly enlarges the rotational
symmetry group. We explicitly construct a nontrivial irreducible Killing tensor associated with the enlarged symmetry group which
permits separation. We also derive first-order equations of motion for particles in these backgrounds and examine some of their properties.

\pagebreak
\setcounter{page}{1}

\tableofcontents
\addtocontents{toc}{\protect\setcounter{tocdepth}{2}}
\newpage

\section{Introduction}

Higher dimensional black hole solutions of the gravitational field equations are currently of great interest due to a large
number of recent developments in high energy physics. These are particularly relevant in the context of superstrings and M-theory, which call for a higher number of
spacetime dimensions (10 or 11). Black hole solutions in these models cannot be ignored, especially when used to describe physics at the Planck scale. Models with large extra dimensions, which
have been proposed in the context of the hierarchy problem and GUT's, naturally include black hole solutions with classical descriptions. Also of interest in these models is the possibility of mini black hole
production in high energy particle colliders which, if they occur, provide a window into non-perturbative gravitational physics.

Astrophysically relevant black hole spacetimes are, to a very good
approximation, described by the Kerr metric \cite{Kerr}. The most natural generalization of
the Kerr metric to higher dimensions, for zero cosmological constant, is given by the Myers-Perry construction
\cite{MyersPerry}. (For a recent generalization with a cosmological constant, see \cite{GPLP}, but a nonzero cosmological constant seems to thwart the type of separability demonstrated in the present paper, so here we shall take the cosmological constant to be zero.) The Myers-Perry metric also does not have charge, but since charged black holes are unlikely to occur in nature, we expect the Myers-Perry type black holes to be the most
 relevant type in spacetimes with extra dimensions.

In this paper, we analyze the separability of the Hamilton-Jacobi equation in Myers-Perry black hole backgrounds in all dimensions. We explicitly perform the separation in the
case where there are only two sets of equal rotation parameters describing the black hole. We use this explicit separation to obtain first-order equations of motion for both massive and massless particles in these backgrounds. The equations are obtained in a form
 that could be used for numerical study.

We study the Klein-Gordon equation describing the propagation of a massive scalar field in this spacetime. Separation is again explicitly shown for the case of two sets of equal black hole rotation parameters. We
construct the separation of both equations explicitly in these cases. We also
construct Killing vectors, which exist due to the additional symmetry, and which permit the separation of these equations.

\section{Overview of the Myers-Perry Metrics}

The Myers-Perry metrics are vacuum solutions of Einstein's equations describing general rotating black hole spacetimes. The Kerr black hole in four dimensions needs an axis of rotation specified. In higher dimensions, this specification is no longer possible. Instead, we
provide rotation parameters specifying rotations in various planes. As such, we use the construction described below.

We introduce $n=[D/2]$ coordinates $\mu_i$ subject to the constraint
\be
\sum_{i=1}^n \mu_i^2 =1\,,
\label{constraint}
\ee
together with $N=[(D-1)/2]$ azimuthal angular coordinates $\phi_i$,
the radial coordinate $r$, and the time coordinate $\tau$.  When the
total spacetime dimension $D$ is odd, $D=2n+1=2N+1$, there are $N=n$
azimuthal coordinates $\phi_i$, each with period $2\pi$.  If $D$ is
even, $D=2n=2N+2$, there are only $N=n-1$ azimuthal coordinates
$\phi_i$. Define $\epsilon$ to be 1 for even $D$, and 0 for odd $D$, so $N=n-\epsilon$.

In Boyer-Lindquist coordinates in $D$ dimensions, the Myers-Perry metrics
are given by

\bea
ds^2 &=& -d\tau^2 + \fft{U\, dr^2}{V-2M} +
\fft{2M}{U}\left(d\tau - \sum_{i=1}^{n-\epsilon} a_i\, \mu_i^2\,
d\phi_i\right)^2 \nn\\
&&+ \sum_{i=1}^n \left(r^2 + a_i^2\right)\, d\mu_i^2 + \sum_{i=1}^{n-\epsilon} \left(r^2 + a_i^2\right)\,
\mu_i^2\, d\phi_i^2 \, , \label{myersp}
\eea
where
\bea
U &=& r^{\epsilon}\, \sum_{i=1}^n \fft{\mu_i^2}{r^2 + a_i^2}\,
\prod_{j=1}^{n-\epsilon} (r^2 + a_j^2)\,, \nonumber \\
F&=& r^2\,  \sum_{i=1}^n \fft{\mu_i^2}{r^2+a_i^2}\,, \nonumber \\
V&=&  r^{\epsilon -2}\prod_{i=1}^{n-\epsilon} (r^2 + a_i^2)
= \fft{U}{F}\,.\label{UFVdef}
\eea
Note that obviously $a_n=0$ in the even dimensional case, as there is no rotation
associated with the last direction.

Since that the metric is block diagonal in the $(\mu _i)$ and the $(r,\tau
,\phi _i)$ sectors, these sectors can be inverted separately. To deal with the $(r,\tau ,\phi _i)$ sector, the most efficient method is to
use the Kerr-Schild construction of the metric. For details on construction of the inverse metric using the Kerr-Schild form, see \cite{VSP}.

We get the following components for the $(r,\tau ,\phi _i)$ sector of $g^{\mu
\nu}$:
\begin{eqnarray}
g^{\tau r}&=&g^{\phi _i r}=0 \,, \nonumber \\
g^{rr}&=&\frac{V-2M}{U}\,, \nonumber \\
g^{\tau \tau}&=&-1-\frac{2MV}{U(V-2M)}\,, \nonumber \\
g^{\tau \phi _i}&=&-\frac{2MVa_i}{U(V-2M)(r^2+a_i ^2)} \,, \nonumber \\
g^{\phi _i \phi_j}&=& \frac{1}{(r^2+a_i ^2)\mu _i ^2}
\delta ^{ij}-\frac{2MV a_i a_j}{U(V-2M)(r^2+a_i ^2)(r^2 +a_j ^2)}  \,, \label{rtfinv}
\end{eqnarray}

Note that the function $U$ depends explicitly on the $\mu_i$'s. Unless
the $(r,\tau,\phi_i)$ sector can be decoupled
from the $\mu$ sector, complete separation is unlikely. If however, all the
$a_i=a$ for some non-zero value $a$, then the $U$ are no longer $\mu$ dependent (taking the constraint into account) and separation seems likely. Note, however, that in this case we cannot deal with even dimensional spacetimes, since $a_n=0$ is different from the other $a_i=a$.

We will actually work with a much more general case, in which separation works in both even and odd dimensional spacetimes. We consider the situation in which the set of rotation parameters $a_i$ take on at most only two distinct values $a$ and $b$ ($a=b$ can be obtained as a special case).
In even dimensions at least one of these values must be zero, since $a_n=0$. As such in even dimensions we take $b=0$ and $a$ to be some (possibly different) value. In the odd dimensional case, there are no restrictions on the values of $a$ and $b$. We adopt the convention
\begin{eqnarray}
a_i = a \qquad \textrm{for} \quad i=1,...,m \quad, \qquad b_j = b \qquad \textrm{for} \quad j=1,...,p \, ,
\end{eqnarray}
where $m+p=N+\epsilon=n$.

Since the $\mu_i$'s are constrained by (\ref{constraint}), we need to use
suitable independent coordinates instead. We use the following decomposition of the $\mu_i$:
\begin{equation}
\mu_i = \lambda _i \sin \theta \quad \textrm{for} \quad i=1,...,m \quad, \qquad \mu_{j+m} = \nu _j \cos\theta \quad \textrm{for} \quad j=1,...,p \, , \label{mudef}
\end{equation}
where the $\lambda_i$ and $\nu_j$ have to satisfy the constraints
\begin{equation}
\sum_{i=1}^{m} \lambda _i ^2 =1 \qquad , \qquad \sum_{j=1}^{p} \nu _ j ^2 =1 \, .
\end{equation}

Since these constraints describe unit $(m-1)$ and $(p-1)$ dimensional spheres in the $\lambda$ and $\nu$ spaces respectively, the natural choice is to use two sets of spherical polar
coordinates. We write
\begin{eqnarray}
\lambda_i&=& \left( \prod _{k=1}^{m-i} \sin\alpha _k\right) \cos \alpha _{m-i+1} \,, \nonumber \\
\nu_j&=& \left( \prod _{k=1}^{p-j} \sin\beta _k\right) \cos \beta _{p-j+1} \,,
\label{lnsphere}
\end{eqnarray}
with the understanding that the products are one when $i=m$ or $j=p$ respectively, and that $\alpha
_m=0$ and $\beta _p =0$.

The $\mu$ sector metric can then be written as
\begin{eqnarray}
ds_{\mu}^2&=&\rho ^2 d\theta ^2 + (r^2+a^2) \sin ^2 \theta \sum_{i=1}^{m-1} \left( \prod _{k=1} ^{i-1} \sin ^2 \alpha _k \right) d\alpha _i ^2 \nonumber \\
&&\qquad +(r^2+b^2) \cos ^2 \theta \sum_{j=1}^{p-1} \left( \prod _{k=1} ^{j-1} \sin ^2 \beta _k \right) d\beta _j ^2 \, , \label{musector}
\end{eqnarray}
again with the understanding that the products are one when $i=1$ or $j=1$. We use the definition
\begin{equation}
\rho ^2 = r^2 + a^2 \cos ^2 \theta + b^2 \sin ^2 \theta \,.
\end{equation}
This diagonal metric can be easily inverted to give
\begin{eqnarray}
g^{\theta \theta} &=& \frac{1}{\rho ^2} \, , \nonumber \\
g^{\alpha _i \alpha _j} &=& \frac{1}{(r^2+a^2)\sin ^2 \theta} \frac{1}{\left(\prod
_{k=1} ^{i-1} \sin ^2 \alpha _k\right)}\delta _{ij} \, , \qquad i,j=1,...,m \, , \nonumber \\
g^{\beta _i \beta _j} &=& \frac{1}{(r^2+b^2)\cos ^2 \theta} \frac{1}{\left(\prod
_{k=1} ^{i-1} \sin ^2 \beta _k\right)}\delta _{ij} \, , \qquad i,j =1,...,p \,. \label{muinv}
\end{eqnarray}

For the case of two sets of rotation parameters that we consider here, the following symbols will be extremely useful in addition to $\rho ^2$:
\begin{eqnarray}
\Delta &=& V-2M \, , \nonumber \\
\Pi &=& \prod_{i=1} ^{N}(r^2+a_i ^2) = (r^2+a^2)^{m} (r^2+b^2)^{p-\epsilon} \, , \nonumber \\
Z &=& (r^2+a^2)(r^2+b^2) \, . \label{newfns}
\end{eqnarray}
Note that these are functions of the variable $r$ only. We note that $U=\frac{r^{\epsilon}\rho ^2 \Pi}{Z}$.
\section{The Hamilton-Jacobi Equation and Separation}
The Hamilton-Jacobi equation in a curved background is given by
\be
-\frac{\partial S}{\partial l} = H = \frac{1}{2} g^{\mu \nu } \frac{\partial
S}{\partial x^{\mu}} \frac{\partial S}{\partial x^{\nu}} \,, \label{HJ}
\ee
where $S$ is the action associated with the particle and $l$ is some affine
parameter along the worldline of the particle. Note that this treatment also
accommodates the case of massless particles, where the trajectory cannot be
parametrized by proper time.

We can attempt a separation of coordinates as follows. Let
\begin{equation}
S=\frac{1}{2}m^2 \lambda -E\tau + \sum_{i=1} ^m \Phi_i \phi _i +\sum_{i=1}^{p} \Psi_i \phi _{m+i} +S_r (r) + S_{\theta}(\theta)
+\sum _{i=1} ^{m-1} S_{\alpha _i} (\alpha _i) +\sum _{i=1} ^{p-1} S_{\beta _i} (\beta _i)\,. \label{ansatz}
\end{equation}
$\tau$ and $\phi _i$ are cyclic coordinates, so their conjugate momenta are
conserved. The conserved quantity associated with time translation is the energy
$E$, and the conserved quantity associated with rotation in each $\phi _i$ is the corresponding angular
momentum $\Phi_i$ or $\Psi_j$. We also adopt the convention that $\Psi_p=0$ in an even number of spacetime dimensions.

Using (\ref{rtfinv}), (\ref{muinv}), (\ref{newfns}), and (\ref{ansatz}) we write the Hamilton-Jacobi equation (\ref{HJ}) as
\vspace{.5cm}
\begin{eqnarray}
-m^2 = &-& \left( 1+\frac{2MZ}{r^2\rho ^2\Delta}\right) E^2 + \frac{2Ma(r^2+b^2)}{r^2\rho ^2 \Delta} \sum_{i=1}^{m} E\Phi_i + \frac{2Ma(r^2+a^2)}{r^2\rho ^2 \Delta} \sum_{i=1}^{p} E\Psi_i \nonumber \\
&+&\frac{\Delta Z}{r^{\epsilon}\rho ^2 \Pi}\left(\frac{dS_r}{dr}\right) ^2 + \frac{1}{(r^2+a^2)} \sum_{i=1}^m \frac{\Phi _ i ^2}{\mu _i ^2} + \frac{1}{(r^2+a^b)} \sum_{i=1}^p \frac{\Psi _ i ^2}{\mu _{i+m} ^2} \nonumber \\
&-&\frac{2Ma^2(r^2+b^2)}{\Delta r^2 \rho ^2 (r^2+a^2)} \sum_{i=1}^m \sum_{j=1}^m \Phi _i \Phi _j -\frac{2Mb^2(r^2+a^2)}{\Delta r^2 \rho ^2 (r^2+b^2)} \sum_{i=1}^p \sum_{j=1}^p \Psi _i \Psi _j \nonumber \\
&-&\frac{4Mab}{\Delta r^2 \rho ^2} \sum _{i=1} ^m\sum_{j=1}^p \Phi _i \Psi _j + \sum_{i=1} ^{m-1} \frac{1}{(r^2+a^2)\sin ^2 \theta \prod _{k=1} ^{i-1} \sin ^2 \alpha _k } \left( \frac{dS_{\alpha _i}}{d\alpha _i}\right) ^2 \nonumber \\
&+& \sum_{i=1} ^{p-1} \frac{1}{(r^2+b^2)\cos ^2 \theta \prod _{k=1} ^{i-1} \sin ^2 \beta _k } \left( \frac{dS_{\beta _i}}{d\beta _i}\right) ^2 -\frac{1}{\rho ^2} \left( \frac{dS_{\theta}}{d\theta}\right)^2 \,.
\end{eqnarray}

Note that here the $\mu_i$ are not coordinates, but simply quantities defined by
(\ref{mudef}). We continue to use the convention defined for products of $\sin ^2 \alpha _i$ and $\sin ^2 \beta _j$ defined earlier.
Separate the $\alpha_i$ and $\beta _j$ coordinates from the Hamilton-Jacobi equation via
\vspace{.5cm}
\begin{eqnarray}
J_1 ^2&=&\sum_{i=1}^m \left[ \frac{\Phi _i ^2}{\lambda _i ^2} + \frac{1}{\prod _{k=1} ^{i-1} \sin ^2 \alpha _k}\left( \frac{dS_{\alpha _i}}{d\alpha _i}\right) ^2\right] \, ,\nonumber \\
L_1 ^2&=&\sum_{i=1}^p \left[ \frac{\Psi _i ^2}{\nu _i ^2} + \frac{1}{\prod _{k=1} ^{i-1} \sin ^2 \beta _k}\left( \frac{dS_{\beta _i}}{d\beta _i}\right) ^2\right] \, , \label{albesep}
\end{eqnarray}
where $J_1^2$ and $L_1^2$ are separation constants. Then the remaining terms in the Hamilton-Jacobi equations can be explicitly separated to give ordinary differential equations for $r$ and $\theta$:
\vspace{1cm}
\begin{eqnarray}
K=&-&m^2r^2+E^2\left(r^2+\frac{2MZ}{r^2\Delta}\right) - \frac{\Delta Z}{r^{\epsilon}\Pi}\left(\frac{dS_r}{dr}\right)^2 - \frac{2Ma(r^2+b^2)}{r^2\Delta} \sum_{i=1} ^ m E \Phi_i \nonumber \\
&-& \frac{2Mb(r^2+a^2)}{r^2\Delta} \sum_{i=1} ^ {p} E \Psi_i + \frac{2Ma^2(r^2+b^2)}{\Delta r^2(r^2+a^2)}\sum_{i=1}^m \sum_{j=1} ^m \Phi _i \Phi _j  +\frac{4Mab}{\Delta r^2} \sum_{i=1}^m \sum_{j=1}^{p} \Phi _i \Psi _j \nonumber \\
&+& \frac{2Mb^2(r^2+a^2)}{\Delta r^2(r^2+b^2)}\sum_{i=1}^p \sum_{j=1} ^p \Psi _i \Psi _j - \frac{r^2+b^2}{r^2+a^2}J_1^2 - \frac{r^2+a^2}{r^2+b^2}L_1^2 \, , \label{rsep} \\
K=&&(m^2-E^2)(a^2\cos ^2 \theta + b^2\sin ^2 \theta) + \left(\frac{dS_{\theta}}{d\theta}\right)^2 + \cot ^2 \theta J_1 ^2 + \tan ^2 \theta L_1 ^2 \label{thetsep}\, ,
\end{eqnarray}
where $K$ is a separation constant.

In order to show complete separation of the Hamilton-Jacobi equation, we analyze the $\alpha$ and $\beta$ sectors in (\ref{albesep}) and demonstrate separation of the individual $\alpha _i$ and $\beta _j$. The pattern here is that of a Hamiltonian of classical (non-relativistic)
particles on the unit $(m-1)$-$\alpha$ and the unit $(p-1)$-$\beta$ spheres, with some potential dependent on the squares of the
$\mu_i$. This can easily be additively separated following the usual procedure, one angle at a time, and the pattern continues for all integers $m,p \ge 2$.

The separation has the following inductive form for $k=1,...,m-2$, and $l=1,...,p-2$:
\begin{eqnarray}
\left(\frac{dS_{\alpha _k}}{d\alpha _k}\right)^2 &=& J_k^2 - \frac{J^2_{k+1}}{\sin ^2 \alpha _k} - \frac{\Phi ^2 _{m-k+1}}{\cos ^2 \alpha _k} \, , \nonumber \\
\left(\frac{dS_{\alpha _{m-1}}}{d\alpha _{m-1}}\right)^2 &=& J_{m-1}^2 - \frac{\Phi_1^2}{\sin ^2 \alpha _{m-1}} - \frac{\Phi ^2 _2}{\cos ^2 \alpha _{m-1}} \, , \nonumber \\
\left(\frac{dS_{\beta _l}}{d\beta _l}\right)^2 &=& L_l^2 - \frac{L^2_{l+1}}{\sin ^2 \beta _l} - \frac{\Psi ^2 _{p-l+1}}{\cos ^2 \beta _l} \, , \nonumber \\
\left(\frac{dS_{\beta _{p-1}}}{d\beta _{p-1}}\right)^2 &=& L_{p-1}^2 - \frac{\Psi_1^2}{\sin ^2 \beta _{p-1}} - \frac{\Psi ^2 _2}{\cos ^2 \beta _{p-1}} \, . \label{albesep2}
\end{eqnarray}

Thus, the Hamilton-Jacobi equation in the Myers-Perry rotating black hole background with
two sets of possibly unequal rotation parameters has the general separation
\begin{equation}
S=\frac{1}{2}m^2 \lambda -E\tau + \sum_{i=1} ^m  \Phi_i \phi _i + \sum_{i=1}^{p}\Psi_i \phi _{m+i}+S_r (r) +S_{\theta}(\theta)+\sum _{i=1} ^{m-1}
S_{\alpha _i}(\alpha _i) +\sum _{i=1} ^{p-1}S_{\beta _i}(\beta _i)\,,
\end{equation}
where the $\alpha _i$  and $\beta _j$ are the spherical polar coordinates on the unit $(m-1)$ and unit $(p-1)$
spheres respectively. $S_r(r)$ can be obtained by quadratures from (\ref{rsep}), $S_{\theta}(\theta)$ by quadratures from (\ref{thetsep}), and the
$S_{\alpha _i}(\alpha _ i)$ and the $S_{\beta_j}(\beta _j)$ again by quadratures from (\ref{albesep2}).

\section{The Equations of Motion}
\subsection{Derivation of the Equations of Motion}

To derive the equations of motion, we will write the separated action $S$ from the Hamilton-Jacobi equation
in the following form:
\bea
S&=&\frac{1}{2}m^2 \lambda -E\tau + \sum_{i=1} ^m \Phi_i \phi _i + \sum _{i=1}^p \Psi_i \phi _i + \int ^r \sqrt{R(r')} dr' + \int ^ {\theta} \sqrt{\Theta(\theta')}d\theta' \nonumber \\
&&\qquad+\sum _{i=1} ^{m-1} \int ^{\alpha _i} \sqrt{A_i (\alpha' _i)}d\alpha ' _i + \sum _{i=1} ^{p-1} \int ^{\beta _i} \sqrt{B_i (\beta' _i)}d\beta ' _i \, ,
\eea
where
\begin{eqnarray}
A_k&=&J_k^2 - \frac{J^2_{k+1}}{\sin ^2 \alpha _k} - \frac{\Phi ^2 _{m-k+1}}{\cos ^2 \alpha _k}\,,\qquad k=1,...,m-2\,, \nonumber \\
A_{m-1}&=&J_{m-1}^2 - \frac{\Phi_1^2}{\sin ^2 \alpha _{m-1}} - \frac{\Phi ^2 _2}{\cos ^2 \alpha _{m-1}} \, , \nonumber \\
B_k&=&L_k^2 - \frac{L^2_{k+1}}{\sin ^2 \beta _k} - \frac{\Psi ^2 _{p-k+1}}{\cos ^2 \beta _k}\,,\qquad k=1,...,p-2\,, \nonumber \\
B_{p-1}&=&L_{p-1}^2 - \frac{\Psi_1^2}{\sin ^2 \beta _{p-1}} - \frac{\Psi ^2 _2}{\cos ^2 \beta _{p-1}} \, ,
\end{eqnarray}
$\Theta$ is obtained from (\ref{thetsep}) as
\begin{eqnarray}
\Theta=K+(E^2-m^2)(a^2\cos ^2 \theta + b^2\sin ^2 \theta) - \cot ^2 \theta J_1 ^2 - \tan ^2 \theta L_1 ^2 \, ,
\end{eqnarray}
and $R$ is obtained from (\ref{rsep}) as
\begin{eqnarray}
\frac{\Delta Z}{\Pi r^{\epsilon}}R&=&(E^2-m^2)r^2+\frac{2MZ}{r^2\Delta}E^2 - \frac{2Ma(r^2+b^2)}{r^2\Delta} \sum_{i=1} ^ m E \Phi_i \nonumber \\
&-& \frac{2Mb(r^2+a^2)}{r^2\Delta} \sum_{i=1} ^ {p} E \Psi_i + \frac{2Ma^2(r^2+b^2)}{\Delta r^2(r^2+a^2)}\sum_{i=1}^m \sum_{j=1} ^m \Phi _i \Phi _j  +\frac{4Mab}{\Delta r^2} \sum_{i=1}^m \sum_{j=1}^{p} \Phi _i \Psi _j \nonumber \\
&+& \frac{2Mb^2(r^2+a^2)}{\Delta r^2(r^2+b^2)}\sum_{i=1}^p \sum_{j=1} ^p \Psi _i \Psi _j - \frac{r^2+b^2}{r^2+a^2}J_1^2 - \frac{r^2+a^2}{r^2+b^2}L_1^2 \, .
\end{eqnarray}

To obtain the equations of motion, we differentiate $S$ with respect to the
parameters $m^2,K,E,J_i^2,L^2_j,\Phi_i,\Psi_j$ and set these derivatives to equal
other constants of motion. However, we can set all these new constants of motion
to zero (following from freedom in choice of origin for the corresponding
coordinates, or alternatively by changing the constants of integration). Following this procedure, we get the
following equations of motion:
\vspace{.5cm}
\begin{eqnarray}
\frac{\pa S}{\pa m^2}&=&0 \Rightarrow \lambda = \int \frac{\Pi r^{\epsilon +2}}{\Delta Z} \frac{dr}{\sqrt{R}} + \int \frac{(a^2\cos ^2 \theta + b^2 \sin ^2 \theta) d\theta}{\sqrt{\Theta}} \, , \nonumber \\
\frac{\pa S}{\pa K}&=&0 \Rightarrow \int \frac{d\theta}{\sqrt{\Theta}} = \int \frac{\Pi r^{\epsilon}}{\Delta Z} \frac{dr}{\sqrt{R}} \, , \nonumber \\
\frac{\pa S}{\pa J_1^2}&=&0 \Rightarrow \int \frac{d\alpha _1}{\sqrt{A_1}} = \int \frac{\Pi r^{\epsilon}}{\Delta Z} \frac{r^2+b^2}{r^2+a^2} \frac{dr}{\sqrt{R}} + \int \cot ^2 \theta \frac{d\theta}{\sqrt{\Theta}} \, , \nonumber \\
\frac{\pa S}{\pa J_k^2}&=&0 \Rightarrow \int \frac{d\alpha _k}{\sqrt{A_k}} = \int \frac{1}{\sin ^2 \alpha _{k-1}} \frac{d\alpha _{k-1}}{\sqrt{A_{k-1}}} \, , \qquad k=2,...,m-2 \, , \nonumber \\
\frac{\pa S}{\pa L_1^2}&=&0 \Rightarrow \int \frac{d\beta _1}{\sqrt{B_1}} = \int \frac{\Pi r^{\epsilon}}{\Delta Z} \frac{r^2+a^2}{r^2+b^2} \frac{dr}{\sqrt{R}} + \int \tan ^2 \theta \frac{d\theta}{\sqrt{\Theta}} \, , \nonumber \\
\frac{\pa S}{\pa L_l^2}&=&0 \Rightarrow \int \frac{d\beta _l}{\sqrt{B_l}} = \int \frac{1}{\sin ^2 \alpha _{k-1}} \frac{d\beta _{l-1}}{\sqrt{B_{l-1}}} \, , \qquad l=2,...,p-2 \,. \label{inteqs}
\end{eqnarray}
We can obtain $N$ more equations of motion for the variables $\phi$ by differentiating $S$ with respect to the angular momenta $\Phi_i$ and $\Psi_j$. Another equation can also be obtained by differentiating $S$ with respect to $E$ involving the time coordinate $\tau$. However, these equations are not particularly illuminating, but can be written out
explicitly if necessary by following this procedure. It is often more convenient to rewrite these in the form of first-order differential equations obtained from (\ref{inteqs}) by direct differentiation with respect to the affine parameter. We only list the most relevant ones here:
\begin{eqnarray}
\rho ^2 \frac{dr}{d\lambda} &=& \frac{\Delta Z}{\Pi r^{\epsilon}} \sqrt{R} \, , \nonumber \\
\rho ^2 \frac{d\theta}{d\lambda} &=& \sqrt{\Theta} \, , \nonumber \\
(r^2+a^2) \frac{d\alpha _k}{d\lambda} &=& \frac{\sqrt{A_k}}{\sin ^2 \theta \prod _{i=1}^{k-1} \sin ^2 \alpha _i} \, , \qquad k=1,...,m-1 \, ,\nonumber \\
(r^2+b^2) \frac{d\beta _k}{d\lambda} &=& \frac{\sqrt{B_l}}{\cos ^2 \theta \prod _{i=1}^{l-1} \sin ^2 \beta _i} \, , \qquad l=1,...,p-1 \, , \label{eqns}
\end{eqnarray}

\subsection{Analysis of the Radial Equation}
The worldline of particles in the Myers-Perry black hole backgrounds considered above
are completely specified by the values of the conserved quantities $E,\Phi _i,\Psi _j, J_i^2, L_j^2$, and by the initial values of the coordinates. We will consider
particle motion in the black hole exterior. Allowed regions of particle motion
necessarily need to have positive value for the quantity $R$, owing to equation
(\ref{eqns}). At large $r$, the dominant contribution to $R$ is
$E^2 -m^2$. Thus we can say that for $E^2<m^2$, we cannot have unbounded orbits,
whereas for $E^2>m^2$, such orbits are possible.

In order to study the radial motion of particles in these metrics, it is useful
to cast the radial equation of motion into a different form. Decompose $R$ as a
quadratic in $E$ as follows:
\begin{equation}
R=\alpha E^2-2\beta E + \gamma \,,
\end{equation}
where
\begin{eqnarray}
\alpha &=& \frac{\Pi r^{\epsilon}}{\Delta Z} \left(r^2+\frac{2MZ}{r^2\Delta}\right) \, , \nonumber \\
\beta &=& -\frac{2M\Pi r^{\epsilon-2}}{\Delta^2 Z}\left( a(r^2+b^2) \sum _{i=1}^m E\Phi _i + b(r^2+a^2) \sum _{i=1}^p E\Psi _i \right) \,, \nonumber \\
\gamma &=& -\frac{\Pi r^{\epsilon}}{\Delta R} \left[ \frac{2M}{r^2\Delta} \left( a^2(r^2+b^2)\sum_{i=1}^m \sum _{j=1}^m \Phi _i \Phi _j + 2ab\sum_{i=1}^m \sum_{j=1}^p \Phi _i \Psi _j \right.\right. \nonumber \\
&& \left.\left.\qquad +b^2(r^2+a^2)\sum_{i=1}^p \sum_{i=1}^p \Psi_i \Psi_j \right)  -m^2r^2 -\frac{r^2+b^2}{r^2+a^2}J_1^2 -\frac{r^2+a^2}{r^2+b^2}L_1^2\right] \, ,
\end{eqnarray}

The turning points for trajectories in the radial motion (defined by the condition
$R=0$) are given by $E=V_{\pm}$ where
\be
V_{\pm} = \frac{\beta \pm \sqrt{\beta ^2 -\alpha \gamma }}{\alpha} \,.
\ee
These functions, called the effective potentials \cite{frolov1},
determine allowed regions of motion. In this form, the radial equation
is much more suitable for detailed numerical analysis for specific values of
parameters.

\subsection{Analysis of the Angular Equations}
Another class of interesting motions possible describes motion at a constant
value of $\alpha _i$ or $\beta_j$. These motions are described by the simultaneous equations
\begin{equation}
A_i (\alpha_i= \alpha _{i0})=\frac{dA _i}{d\alpha _i}(\alpha _i=\alpha _{i0}) =0 \,, \qquad i=1,...,m-1\,,
\end{equation}
in the case of constant $\alpha_i$ motion, where $\alpha _{i0}$ is the constant value of $\alpha _i$ along this trajectory, or by the simultaneous equations
\begin{equation}
B_i (\beta_i= \beta _{i0})=\frac{dB _i}{d\beta _i}(\beta _i=\beta _{i0}) =0 \,, \qquad i=1,...,p-1\,,
\end{equation}
in the case of constant $\beta_i$ motion, where $\beta _{i0}$ is the constant value of $\beta _i$ along this trajectory.

These equations can be explicitly solved. In the case of constant $\alpha _i$ motion, we get the relations
\begin{eqnarray}
\frac{J^2_{i+1}}{\sin ^4 \alpha _i} &=& \frac{\Phi^2 _{m-i-1}}{\cos ^4 \alpha _i} \,,\nonumber \\
J^2_i&=&\frac{J^2_{i+1}}{\sin ^2 \alpha _i} +\frac{\Phi^2_{m-i+1}}{\cos ^2 \alpha_i}\,,\qquad i=1,...,m-1\,.
\end{eqnarray}
Note that if $\alpha_{i0}=0$, then $J^2 _{i+1} =0$, and if $\alpha _{i0} = \pi /2$, then $\Phi_{m-i+1} ^2=0$. Similarly, in the case of constant $\beta_i$ motion, we get the relations
\begin{eqnarray}
\frac{L^2_{i+1}}{\sin ^4 \beta _i} &=& \frac{\Psi^2 _{p-i-1}}{\cos ^4 \beta _i} \,,\nonumber \\
L^2_i&=&\frac{L^2_{i+1}}{\sin ^2 \beta _i} +\frac{\Psi^2_{p-i+1}}{\cos ^2 \beta_i}\,,\qquad i=1,...,p-1\,.
\end{eqnarray}
Again if $\beta_{i0}=0$, then $L^2 _{i+1} =0$, and if $\beta _{i0} = \pi /2$, then $\Psi_{p-i+1} ^2=0$.

Examining $A_k $ in the general case, $\alpha _k=0$ can only be reached if $J_{k+1}=0$, and
$\alpha _k =\pi/2$ can be only be reached if $\Phi_{m-k+1}=0$. The orbit will completely be in the subspace
$\alpha _ k=0$ only if $J_k ^2 = \Phi^2 _{m-i+1}$ and will completely be in the subspace $\alpha _k =\pi/2$ only if $J_k ^2 = J_{k+1} ^2$. Analogous results hold for constant $\beta _i$ motion.

Again these equations are in a form suitable for numerical analysis for specific values of the black hole and particle parameters.

\section{Dynamical Symmetry}
The general class of metrics discussed here are stationary and ``axisymmetric";
i.e., $\partial / \partial \tau$ and $\partial / \partial \varphi _i$ are Killing
vectors and have associated conserved
quantities, $-E$ and $L_i$. In general if $\xi$ is a Killing vector, then $\xi
^{\mu} p_{\mu}$ is a conserved quantity, where $p$ is the momentum. Note that this quantity is first order in the momenta.

With the assumption of only two sets of possibly unequal rotation parameters, the
spacetime acquires additional dynamical symmetry and more Killing vectors are
generated. We have complete symmetry between the various planes of rotation characterized by the same value of rotation parameter $a_i=a$,
and we can ``rotate" one into another. Similarly, we have symmetry between the planes of rotation characterized by the same value of the rotation parameter $a_i=b$, and we can ``rotate" these into one another as well. The vectors that generate these
transformations are the required Killing vectors. The explicit construction of such Killing vectors is done in \cite{VSP}. In this case, we get two independent sets of such Killing vectors, associated with the constant $a$ and $b$ value rotations.

These Killing vectors exist since the rotational symmetry of the spacetime has been greatly enhanced. In  an odd number of spacetime dimensions, if $a\neq b$ and both are nonvanishing, then the rotational symmetry group is $U(m)\times U(p)$. If one of them is zero, but the other is nonzero (we take the nonzero one to be $a$), then the rotational symmetry group is $U(m)\times O(2p)$. In the case when $a=b\neq0$, the rotational symmetry group is
$U(m+p)$. In the case when $a=b=0$, i.e. in the Schwarzschild metric, the rotational symmetry group is $O(2m+2p)$. In an even number of spacetime dimensions, $b=0$ in the cases we have analyzed. If $a\neq0$, then the rotational symmetry group is $U(m)\times O(2p-1)$, and in the case when $a=b=0$, i.e. in the Schwarzschild metric, the rotational
symmetry group is $O(2m+2p-1)$. Note that since these metrics are stationary, the full dynamical symmetry group is the direct product of $\mathbf{R}$ and the rotational symmetry group, where $\mathbf{R}$ is the additive group of real numbers parametrizing $\tau$.

In addition to these reducible angular Killing tensors, we also obtain a non-trivial irreducible second-order Killing tensor, which permits the separation of the $r-\theta$ equations. This Killing tensor is a generalization of the result obtained in the five dimensional case in \cite{frolov1}. This is obtained from the separation constant $K$ in (\ref{rsep}) and (\ref{thetsep}).
We choose to analyze the latter.
\begin{equation}
K=(m^2-E^2)(a^2\cos ^2 \theta + b^2 \sin ^2\theta) + \cot ^2 \theta J_1 ^2 + \tan ^2 \theta L_1 ^2 + \left(\frac{\partial S}{\partial \theta}\right)^2 \, .
\end{equation}
The Killing tensor $K^{\mu \nu}$ is obtained from this separation constant (which is quadratic in the canonical momenta) using the relation $K=K^{\mu \nu} p_{\mu} p_{\nu}$. Its is then easy to see that
\begin{equation}
K^{\mu \nu} = \left(g^{\mu\nu}-\delta ^{\mu}_{\tau}\delta ^{\nu}_{\tau}\right)\left(a^2\cos ^2 \theta + b^2 \sin ^2 \theta \right) + \cot ^2 \theta J_1 ^{\mu \nu} + \tan ^2 \theta L_1 ^{\mu \nu} + \delta ^{\mu}_{\theta}\delta ^{\mu}_{\theta} \, ,
\end{equation}
where $J_1^{\mu \nu}$ and $L_1^{\mu \nu}$ are the reducible Killing tensors associated with the $\alpha_1$ and $\beta_1$ separation.

It is the existence of these additional Killing vectors and the nontrivial irreducible Killing tensor, due to the increased symmetry of the spacetime, which permits complete separation of the Hamilton-Jacobi equation.

\section{The Scalar Field Equation}
Consider a scalar field $\Psi$ in a gravitational background with the action
\begin{equation}
S[\Psi]=-\frac{1}{2}\int d^Dx\sqrt{-g}((\nabla \Psi)^2+ \alpha R \Psi ^2 + m^2
\Psi ^2 ) \,,
\end{equation}
where we have included a curvature-dependent coupling. However, the
Myers-Perry background is Ricci flat since it is a solution to the vacuum Einstein equations, so $R=0$.
Variation of the action leads to the Klein-Gordon equation
\begin{equation}
\frac{1}{\sqrt{-g}}\partial _{\mu}(\sqrt{-g} g^{\mu \nu}\partial _{\nu} \Psi
)=m^2 \Psi \,.\label{KG1}
\end{equation}

As discussed by Carter \cite{Carter}, the assumption of
separability of the Klein-Gordon equation usually
implies separability of the Hamilton-Jacobi equation. Conversely, if the
Hamilton-Jacobi equation does not separate, the Klein-Gordon equation seems
unlikely to separate. We can also see this explicitly (as in the case of the Hamilton-Jacobi equation), since the $(r,\tau,\phi_i)$ sector
has coefficients in the equations that explicitly depend on the $\mu_i$ except when of all $a_i=a$, in which case separation seems likely. We will again consider the much more general case of
two sets of possibly unequal sets of rotation parameters $a$ and $b$. We continue using the same numbering conventions for the variables.

Once again, we impose the constraint (\ref{constraint}) and decompose the
$\mu_i$ in two sets of spherical polar coordinates as in (\ref{mudef}) and (\ref{lnsphere}). We calculate
the determinant of the metric to be
\begin{eqnarray}
g&=&-r^2 \rho ^4 \Pi (r^2+a^2)^{m-2}(r^2+b^2)^{p-2}\sin ^{4m-2}\theta \cos ^{4p-2-2\epsilon}\theta \left[ \prod _{j=1}^{m-1} \sin ^{4m-4j-2}\alpha _j \cos ^2 \alpha _j \right]  \nonumber \\
&&\, \left[ \prod _{k=1}^{p-1} \sin ^{4p-4k-2}\beta _k \cos ^2 \beta _k \right] \cos ^{-2\epsilon} \beta _1 \,.
\end{eqnarray}
For convenience we write $g=-RTAB\rho ^4$, where
\begin{eqnarray}
R&=&r^2\Pi(r^2+a^2)^{m-2}(r^2+b^2)^{p-2} \, ,\nonumber \\
T&=&\sin^{4m-2}\theta \cos ^{4p-2-2\epsilon} \theta \, ,\nonumber \\
A&=&\prod _{j=1}^{m-1} \sin ^{4m-4j-2}\alpha _j \cos ^2 \alpha _j \, ,\nonumber \\
B&=&\prod _{k=1}^{p-1} \sin ^{4p-4k-2}\beta _k \cos ^2 \beta _k \cos ^{-2\epsilon} \beta _1 \, .
\end{eqnarray}
Note that $R$ and $T$ are functions of $r$ and $\theta$ only, and $A$ and $B$ only depend on the set of variables $\alpha _i$ and $\beta _j$ respectively.
Then the Klein-Gordon equation in this background (\ref{KG1}) becomes
\begin{eqnarray}
m^2 \Psi &=& \frac{1}{\rho ^2 \sqrt {R}} \partial _r \left( \sqrt{R} \frac{\Delta Z}{r^{\epsilon}\Pi} \partial _r \Psi \right) -\left[1+\frac{2MZ}{r^2\rho ^2 \Delta}\right] \partial _ {\tau} ^2 \Psi + \frac{1}{r^2+a^2}\sum_{i=1} ^{m} \frac{1}{\mu _i ^2}\partial _{\phi_i} ^2 \Psi \nonumber \\
&+& \frac{1}{r^2+b^2}\sum_{i=1} ^{p-\epsilon} \frac{1}{\mu _{m+i} ^2}\partial _{\phi_{i+m}} ^2 \Psi   -\frac{2Ma^2(r^2+b^2)}{\Delta r^2 \rho ^2 (r^2+a^2)} \sum _{i=1}^m \sum _{j=1} ^m \partial _{\phi _i} \partial _{\phi _j} \Psi \nonumber \\
&-& \frac{2M}{r^2\rho ^2 \Delta}\left[ a(r^2+b^2)\sum _{i=1}^m \partial _{\tau} \partial _{\phi _i} \Psi + b(r^2+a^2)\sum _{i=1}^{p-\epsilon} \partial _{\tau} \partial _{\phi _{m+i}} \Psi\right]  \nonumber \\
&-& \frac{2Mb^2(r^2+a^2)}{\Delta r^2 \rho ^2 (r^2+b^2)} \sum _{i=1}^{p-\epsilon} \sum _{j=1} ^{p-\epsilon} \partial _{\phi _{i+m}} \partial _{\phi _{j+m}} \Psi-\frac{4Mab}{\Delta r^2 \rho ^2 } \sum _{i=1}^m \sum _{j=1} ^{p-\epsilon} \partial _{\phi _i} \partial _{\phi _{j+m}} \Psi  \nonumber \\
&+& \frac{1}{\rho ^2 \sqrt {T}} \partial _{\theta} \left( \sqrt{T} \partial _ {\theta} \Psi\right)+ \frac{1}{(r^2+a^2)\sin ^2 \theta} \left[\sum _{i=1} ^{m-1} \frac{1}{\sqrt{A}} \partial _{\alpha_i} \left( \frac{\sqrt{A}}{\prod _{k=1} ^ {i-1} \sin ^2 \alpha _k } \partial _{\alpha _i} \Psi\right)\right] \nonumber \\
&+&\frac{1}{(r^2+b^2)\cos ^2 \theta} \left[\sum _{i=1} ^{p-1} \frac{1}{\sqrt{B}} \partial _{\beta_i} \left( \frac{\sqrt{B}}{\prod _{k=1} ^ {i-1} \sin ^2 \beta _k } \partial _{\beta _i} \Psi \right)\right] \,.
\end{eqnarray}
We attempt the usual multiplicative separation for $\Psi$ in the following form:
\begin{equation}
\Psi=\Phi_r (r) \Phi_{\theta}(\theta)e^{-iEt}e^{i\sum_i^m \Phi_i \phi _i} e^{i\sum_i^{p-\epsilon} \Psi_i \phi _{m+i}}\left(\prod_{i=1}^{m-1}\Phi_{\alpha_i}(\alpha_i)\right)\left(\prod_{i=1}^{m-1}\Phi_{\beta_i}(\beta_i)\right)\,. \label{KGanz}
\end{equation}
The Klein-Gordon equation then completely separates. The $r$ and $\theta$ equations are given as
\begin{eqnarray}
K&=&\frac{1}{\Phi_r \sqrt{R}} \frac{d}{dr}\left(\sqrt{R} \frac{\Delta Z}{r^{\epsilon}\Pi} \frac{d\Phi _r}{dr}\right) + r^2(E^2-m^2) +\frac{2MZE^2}{r^2 \Delta} \nonumber \\
&-& \frac{2ME}{r^2\Delta}\left[ a(r^2+b^2) \sum _{i=1}^m \Phi_i + b(r^2+a^2) \sum _{i=1}^{p-1\epsilon} \Psi_i \right] \nonumber \\
&+&\frac{2M}{\Delta r^2} \left[ \frac{a^2(r^2+b^2)}{(r^2+a^2)}\sum _{i=1} ^m \sum _{j=1} ^m \Phi _i \Phi _j+\frac{b^2(r^2+a^2)}{(r^2+b^2)}\sum _{i=1} ^{p-\epsilon} \sum _{j=1} ^{p-1\epsilon} \Psi _i \Psi _j + 2ab\sum _{i=1}^m \sum _{j=1} ^{p-\epsilon} \Phi _i \Psi_j\right] \, ,\nonumber \\
-K&=& \frac{1}{\Phi_{\theta}\sqrt{T}}\frac{d}{d\theta}\left(\sqrt{T} \frac{d\Phi_{\theta}}{d\theta}\right) + (E^2-m^2)(a^2\cos ^2 \theta + b^2 \sin ^2 \theta) \nonumber \\
&+& K_1 \cot ^2 \theta + M_1 \tan ^2 \theta \,,
\end{eqnarray}
where $K$, $K_1$ and $M_1$ are separation constants. $K_1$ and $M_1$ encode all the $\alpha$ and $\beta$ dependence respectively and are defined explicitly as follows:
\begin{eqnarray}
K_1 &=& \sum _{i=1} ^{k-1} A_i + \frac{ K_k }{\prod _ {j=1} ^{k-1} \sin ^2\alpha _j }\,, \quad k=1,...,m-1\,,
\end{eqnarray}
where
\begin{eqnarray}
A_i &=&  \frac{1}{\Phi _{\alpha _i} \cos \alpha _i \sin
^{2m-2i-1} \alpha _i \prod _{k=1} ^{i-1} \sin ^2 \alpha _ k}
\frac{d}{d\alpha _i} \left( \cos \alpha _i \sin ^{2m-2i-1} \alpha _i \frac{d\Phi
_{\alpha _i}}{d\alpha _i} \right) \nonumber \\ && -\frac{\Phi^2 _{m-i+1}}{\cos ^2 \alpha _i \prod _{j=1} ^{i-1} \sin
^2{\alpha _j}}  \,,
\end{eqnarray}
and
\begin{eqnarray}
M_1 &=& \sum _{i=1} ^{k-1} B_i + \frac{ M_k }{\prod _ {j=1} ^{k-1} \sin ^2\beta _j }\,, \quad k=1,...,p-1\,,
\end{eqnarray}
and where
\begin{eqnarray}
B_i &=&  \frac{1}{\Psi _{\beta _i} \cos \beta _i \sin
^{2p-2i-1} \beta _i \prod _{k=1} ^{i-1} \sin ^2 \beta _ k}
\frac{d}{d\beta _i} \left( \cos \beta _i \sin ^{2p-2i-1} \beta _i \frac{d\Phi
_{\beta _i}}{d\beta _i} \right) \nonumber \\ && -\frac{\Psi^2 _{p-i+1}}{\cos ^2 \beta _i \prod _{j=1} ^{i-1} \sin
^2{\beta _j}}  \,,
\end{eqnarray}
Then we inductively have the complete separation of the $\alpha _i$ dependence as
\begin{equation}
K_k  = \frac{K _{k+1}}{\sin ^2 \alpha _k} -\frac{\Phi^2 _{n-k+1}}{\cos ^2
\alpha _k} + \frac{1}{\Phi _{\alpha_k} \cos \alpha _k \sin ^{2m-2k-1} \alpha_k}
\frac{d}{d\alpha _k} \left( \cos\alpha _k \sin \alpha _k \frac{d\Phi _{\alpha
_k}}{d\alpha _k} \right) \,,
\end{equation}
where $k=1,...,m-1$, and we use the convention $K_m = -\Phi_1 ^2$. Similarly, the complete separation of the $\beta_i$ dependence is given inductively by
\begin{eqnarray}
M_k  = \frac{M _{k+1}}{\sin ^2 \beta _k} -\frac{\Psi^2 _{p-k+1}}{\cos ^2
\beta _k} + \frac{1}{\Phi _{\beta_k} \cos \beta _k \sin ^{2p-2k-1} \beta_k}
\frac{d}{d\beta _k} \left( \cos\beta _k \sin \beta _k \frac{d\Phi _{\beta
_k}}{d\beta _k} \right) \,,
\end{eqnarray}
where $k=1,...,p-1$, and we use the convention $M_p = -\Psi_1 ^2$.
These results agree with the previously known analysis in five dimensions \cite{frolov2}.

At this point we have complete separation of the Klein-Gordon equation in the Myers-Perry black hole background in all dimensions with two sets of possibly unequal rotation parameters in the form given by
(\ref{KGanz}) with the individual separation functions given by the ordinary differential equations above. Note that the separation of the Klein-Gordon equation in this geometry is again
due to the fact that the symmetry of the space has been enlarged.

\section*{Conclusions}
We studied the separability properties
of the Hamilton-Jacobi and the Klein-Gordon equations in the Myers-Perry black hole backgrounds in all dimensions.
Separation in Boyer-Lindquist coordinates is possible for the case of two possibly unequal sets of rotation parameters. This is due to the
enlarged dynamical symmetry of the spacetime. We discuss the Killing vectors and reducible Killing tensors that exist in the spacetime. In addition we construct the nontrivial irreducible Killing tensor which explicitly permits complete separation.
 Thus we demonstrate the separability of the Hamilton-Jacobi and the Klein-Gordon equations as a direct consequence of the enhancement of symmetry.
We also derive first-order equations of motion for classical particles in these backgrounds, and analyze the properties of some special trajectories.

Further work in this direction could include the study of higher-spin field equations in these backgrounds, which is of great interest,
particularly in the context of string theory. Explicit numerical study of the equations of motion for specific values of the black hole parameters
could lead to interesting results.

\section*{Acknowledgments}

We are grateful to Gary Gibbons for providing a copy of earlier work by Rebecca Palmer\cite{Palmer} that made progress toward separation in higher
dimensional Myers-Perry metrics. Our research was supported in part by the Natural Science and Engineering Research Council of
Canada.

\end{document}